\documentclass[11pt]{article}
\usepackage[pdftex]{graphicx}  
\usepackage{graphicx}
\pdfoutput=1
\newcommand{\di}{\rm d}
\newcommand{\bxi}{\mbox{\boldmath$\xi$}}

\begin{document} 
\setcounter{page}{1}

\vspace*{5mm}

\begin{center}
\textbf{\Large{{{The growth and development of living organisms \\ from the thermodynamic point of view}}}}{\Large{{}}}\\

\vspace{5mm}

\textit{\large{{{Alexei A. Zotin} }}}{\large{{}}}\\
{Koltzov' Institute of Developmental Biology RAS 
\\
Vavilov Str. 26, Moscow, Russia 119334 }\\

\vspace{2mm}

\textit{\large{{{and} }}}{\large{{}}}\\
 {\large{{ }}}{\large{{ }}}

\vspace{2mm}
\textit{\large{{{Vladimir N. Pokrovskii} }}}{\large{{}}}\\
 {\large{{ }}}{\large{{ }}}
{Moscow State University of Economics, Statistics and Informatics,
\\
 Moscow, RUSSIA 119501}\\
\end{center}

\begin{abstract}

The living organism is considered as an open system, whereas Prigogine's approach to the thermodynamics of such systems is used. The approach allows one to formulate the law of individual growth and development (ontogenesis) of the living organism, whereas it has taken into account that the development and functioning of the system are occurring under the special internal program. The thermodynamic equation of growth is followed a method of estimation of the specific entropy of organism. The theory is compared with experimental data, whereas estimates of the specific  entropy of some species were done; it shows a reduction of specific entropy in the evolutionary row: yeast - insects - reptiles - birds.

\end{abstract}
\begin{center}
\vspace{5mm}

\par\end{center}


\begin{center}
\vfill{}

\par\end{center}

\begin{center}
October   2018

\par\end{center}

\vspace{5mm}

\newpage 



\section{Introduction}

Many prominent investigators [1 - 4]  approached  the formulation of   phenomenological (thermodynamic) theory of the development,  growth and aging of a living organism (theory of ontogenesis). The theories were based on the analysis of  quantities  characterizing  energy exchange between an organism and the surrounding. Measuring simultaneously both the  incoming  flux of energy and  intensity of heat production, the investigators found [5, 6]  that,  under the  growth, the  former  can surpass the  later. The difference between the  incoming  flux of energy and  intensity of heat production, dubbed as  $\psi_u$-function [5, 6], has been qualified as a part of  energy that is needed for  the construction and growth  of the  organism [7]. This idea was central in our  investigation, and we are going  now to formulate the exact relation between $\psi_u$-function and the rate of mass growth of the living organism.

From the thermodynamic point of view, the living organism ought to be  considered as an open thermodynamic system that exists in a non-equilibrium state [4, 8 - 10], and the desire to understand the general principles that govern the living organisms was one of the important motivations to investigate  open systems. The living organism does not contain anything, apart of chemical compounds, and all processes that are occurring in the system  are connected, in the large degree, with chemical transformations, so that  there were some attempts to present  simple models imitating the behaviour of the living organism. The brusselator, considered by Prigogine with collaborators [11],  can be considered as one of the elementary models that reflects the essential features of the biological organism: an exchange of substances with the environment and existence of stable spatial internal structure within a  limited volume. However, the living organism has extra  peculiarities of the behaviour: unlike the brusselator, all processes are running under internal control, trying to keep  the internal  state of organism unchangeable, which is known as a phenomenon of homeostasis [12]. The control mechanism changes the speed of running  of processes depending on the signals receiving from the outside. At the rise of  outer temperature, for example, the control mechanism commands to increase perspiration, for  the temperature of the organism  not to be increased.  This special congenital program also determines the development and growth of the organism.

Do all of this require  some modification of thermodynamic laws at  application to living systems in comparison with chemical systems of the brusselator type?   In this paper, first of all, we are going  to remind main results of  Prigogene's formulation of thermodynamics of open systems (Section 2).  In Section 3 we apply Prigogene's results to show  how the control apparatus can be introduced into the thermodynamic approach (Section 3.1); this  gives a clue to the  problem of the description of growth and development  of the living organism. Then, we are formulating (Section 3.3)  and  testing (Section 3.4)  the thermodynamic equation of growth of  the living organism. The {\it Discussion and Conclusions}  section  contains a discussion of the resolution of the problem presented herein.

\section{The Change of Entropy in  Prigogine's Formulation}

The essential contribution to the thermodynamics of open systems was brought by Prigogine,  when he and his collabourators  investigated  systems of  chemically reacting substances. Stationary  states of such systems exist due to exchange both particles and energy with the environment. In section 8 of the third chapter of his book [9], Prigogine has specified three contributions to the variation of entropy of an  open system. The simple analysis [13, 14], reproduced in Appendix, demonstrates that the change of entropy $S$ of the system  at the given temperature $T$ can be calculated according to the formula 
\begin{equation}
T\,dS = \Delta Q - \sum_{j} \, \Xi_{j} \,\Delta \xi_j  + \sum_{\alpha}\, \eta_\alpha \, \Delta N_\alpha.
\end{equation}
The first term on the right-hand side of the equation presents a stream of thermal energy into the  system; it can be positive (the stream into the system)   or negative (the stream out of the system).  The last term depicts a stream of energy  coming with the stream of particles of substances  $\Delta N_\alpha$ that can be positive (the stream into the system)   or negative (the stream out of the system), $\eta_\alpha$ is connected with chemical potential of substance labeled $\alpha$.   The  middle term on the right-hand side of  equation (1) presents a change of the stored energy inside the system due to the existing   of internal variables $\xi_i$, so called thermodynamic forces $\Xi_{j}$ are considered positive, so that this term gives a positive contribution into entropy at relaxation of internal variables.  In the case of chemically  reacting substances, which  was investigated  by Prigogine [9], the internal variables $\xi_i$  appear to be measures of incompleteness of chemical reactions, that are the  measures of how much  the considered system with chemical reactions is  out of equilibrium; the thermodynamic forces  $\Xi_{j}$ are affinities of corresponding reactions. The theory was  generalized [13, 14] to  consider any deviation from the equilibrium state as an internal variable, so that we can consider the  set of internal variables $\xi_j$ in equation (1) to be comprised of  the quantities determing not only degrees of completeness of all chemical reactions occurring in the system, but also a structure of the system, gradients of temperature, difference of concentrations of substances and so on. Note that equation (1) presents the balance  of energy and is not restricted  to any approximations.

The sign of the middle term on the right-hand side of equation (1)  is taken  in such a way to  ensure the positive production of entropy at the relaxation of internal variables $\xi_j$, but two processes are running simultaneously. Under  some external influences (the fluxes of heat and substances) the system is passing to a new state, while internal variables could  emerge  and entropy of the system could change. Due to internal processes that cannot be controlled from the outside, the internal variables, as  measures of non-equilibrium of the system, tend to disappear. For small deviations of internal variables $\xi_i$ from their equilibrium values $\xi_i^{(0)}$, the local law of disappearing can be written as relaxation equation for each internal variable
\begin{equation}
\frac{d\xi_i}{dt} = - \frac{1}{\tau_i} \, \left(\xi_i - \xi_i^{(0)} \right),\quad i =1,\,2,\ldots , 
\end{equation} 
where $\tau_i= \tau_i(T, x_1, x_2, \ldots, x_n)$ is a relaxation time of a corresponding internal variable that, being left on its own,  permanently decreases and, eventually, comes close to equilibrium value $\xi_i^0$, which is conveniently  considered to be equal to zero. In this situation, due to the recorded equation (2), the middle term on the right-hand side of equation (1),  is positive and describes the  dissipation of the stored energy.   The dynamics of internal variables for arbitrary deviations from equilibrium was formulated in the paper [14].   

Note  that, in line with the notion of entropy of the open system $S$, one can introduce the  notion of {\it complexity}; the latter has the same sense as notion of {\it bound information} [15]  in application to complex systems. The complexity can be considered as a characteristic of the set of  internal variables $\xi_i$: the internal variables themselves sometimes are called {\it variables of complexity} or {\it variables of order}.   Though at the moment it is impossible to define  how complexity  is related to the number and connectedness of internal variables, one can assert that the greater the number of internal variables,  their values and mutual connections, the greater  the complexity.  In virtue of  equation (1), the change of complexity can be connected  with the change of entropy   of the system:  increase or decrease in number and values of internal variables  provokes  decrease or increase  in entropy.

\section{Organism as an Open Thermodynamic System}

We shall discuss the processes of growth and development  of the living organism, considering it as an open  thermodynamic system of undefined volume, but with the certain mass $M$  and  temperature $T$. To describe a  huge number of  internal processes within the system,  one needs  a set of internal variables $\xi_j$, that, in case of the living organism,  are emerged under the influence of the  streams of substances and heat.  On the basis of fundamental expression for variation of entropy of  an open system (1), the equation for the growth rate of specific entropy $S/M$ of the organism can be written as  
\begin{equation}
\frac{M}{S} \,\frac{\di}{\di t} \left(\frac{S}{M}\right) + \frac{1}{M} \, \frac{\di M}{\di t}  = \frac{1}{T S} \left(\frac{\Delta Q}{\Delta t} - \sum_{i} \, \Xi_{i} \,\frac{\di \xi_i }{\di t} + \sum_{\alpha}\, \eta_\alpha \, \frac{\Delta N_\alpha }{\Delta t} \right).
\end{equation}
It is marked with  the symbol $\Delta$ that the corresponding quantities are objects of direct empirical observation.

The law of conservation of mass determines the equation of growth 

\begin{equation}
\frac{\di M}{\di t}  =  \sum_{\alpha}\, m_\alpha \, \frac{\Delta N_\alpha }{\Delta t},
\end{equation}
where $m_\alpha$ is mass of a particle of substance labeled $\alpha$. Unfortunately, this form of the growth equation is useless practically, so as there are a lot of the processes occurring in the living system, and it is extremely difficult to measure the substances coming in and out the organism [4].

The balance equations (3) and (4) give a basis for thermodynamic interpretation of the phenomena of existence, growth and ageing of a biological organism, but for this purpose, it is necessary to establish, first of all, correspondence of all symbols in these parities to the quantities concerning the living organism.

\subsection{Three groups of internal variables} 

Let's remind, that the internal variables that are present in the middle term  of the  right part of the equation (3) determine the deviation of the system from the state of equilibrium and, in this way, apart of the incompleteness of internal chemical reactions, describe the architectural and functional structure of the system. Considering the organism as an open system, we can note that there is permanent changing  of internal variables, which emerge under fluxes of heat and substances and tend to disappear. The rate of disappearing is different for different internal variables; time of relaxation of some of them are very small as compared  with the lifespan of organism. The other internal variables exist during the total lifetime and even more. The difference of internal variables according to their rates of approach the equilibrium state   determines  their  functional distinctions in the processes of the growth and development of the organism.\footnote{Consideration of an hierarchy of relaxation processes is useful in many applications of non-equilibrium thermodynamics. For example,  Gujrati [16] demonstrated recently that the division of internal variables according to their relaxation times is very helpful  for description of the phenomena of solidification of glasses.} For the further discussion, we are dividing  the set of internal variables of the system into three groups, according to values of relaxation times; further on, the different functional roles of separate groups in the processes of development and growth are described.  

First of all, it is possible to separate a group of internal variables, which  originates from the complexity of the first cell that gives the beginning of the organism. The material carriers of complexity are the molecules of a deoxyribonucleic acid packed in the nuclei of the cells. The copies of internal variables (complexity) of the initial set are present in each cell and remain constant (at some idealization) up to the death of an organism. The number of copies of these variables increases at the proliferation of cells, and during the realization of the program of development. These variables comprise  the foundation of a control mechanism that  supervises  all processes in the organism during its existence and  growth. The organism is developing under the installed plan received by right of succession, whereas some internal variables are suppressed or expressed when the cells are multiplying. 
 A role of internal variables of the first group is  the  conservation of hereditary instructions and the control over the construction of an organism. 

The second group of internal variables describes the emerged structure of the organism in processes of morphogenesis and differentiation, when  the interoperability of the parts of an organism is materializing. For example, conformations of  the biomolecules having albuminous in the basis, change in such a manner that speed of reactions in which they participate, also sharply changes. The greater role is also playing a spatial arrangement of fibers and, in particular, enzymes  [17]; it provides such processes that cannot occur in a random mixture of chemical reagents. The number and values  of such internal variables increase at growth and development of the organism.  The set of variables of the second groups determines some metastable structure (organs and tissues) of the living organism and have the times of  relaxation comparable with the lifespan of the organism, so that they do not contribute into the current dissipation  of energy (production of entropy).

The set of internal variables of the third group emerges in the course of the above-described processes of multiplication of original complexity and construction of the structure of the organism. The processes are based on chemical transformations of substances, occur in non-equilibrium situations  and  are accompanied by excitation of internal variables and their fast relaxation. The set of internal variables of this group includes the variables locating gradients of temperature, differences in  concentration of substances and degrees of completeness of all chemical reactions running in the organism. Internal variables of the third group accompany all existing processes anywhere they were. These internal variables have very small times of relaxation (in comparison with existence of the organism) and determine the  basic contribution to the internal production of entropy. The dissipated energy is coming out in the form of heat.

Note that now we can record an expression for entropy of the  organism as a function of the internal variables, whereas the expansion can be done over  internal variables of the third group $\bxi^{(3)}$ 
$$
S(\bxi^{(1)}, \bxi^{(2)}, \bxi^{(3)}) = S_0(\bxi^{(1)}, \bxi^{(2)}) - \, \frac{1}{2 T} S_{ij}(\bxi^{(2)}) \, \xi^{(3)}_i \xi^{(3)}_j.   
$$
The quantity  $S_0$ is entropy of a dead body (a corpse), the quantity is  a decreasing function of variables of the first group $\bxi^{(1)}$ (carriers of instructions for construction and functioning of an organism) and the second group $\bxi^{(2)}$ (the description of metastable  structures of the organism). 
The last term on the right hand side of the equation presents all processes running in the live organism; the quantities  $S_{ij}$ are positive and essentially depend on internal variables of the second group, which define metastable structure (a set  of structures) of the  system and cannot be considered as small quantities. 

The difference in the relaxation times of internal variables allows us to distinguish different functional roles of internal variables in the processes of development and growth. The distribution  of the all  internal variables over three groups can be helpful for designing of the functional scheme of the living organism.

\subsection{Exchange of energy between the organism and environment}

There are methods of assessments of incoming and outgoing fluxes of energy for the  organism of an animal [4]; these fluxes are presented in the right part of the equation (3) by the first and the last terms. The greatest amount of energy is delivered into organisms with food in the form of energy of chemical links of organic molecules [4, 18]. The speed of synthesis of adenosine triphosphate (ATP)  gives us an estimate of the coming-with-food energy (in a unit of time). The synthesis  can be realized in the two basic processes: the  oxidizing phosphorylation and glycolysis. One can assume that the energy received in the process of glycolysis, can be neglected in comparison with the process of oxidizing phosphorylation [4], and  one can judge about the speed of synthesis ATP indirectly, namely on the rate of oxygen consumption or the rate of allocation of carbonic gas. It is shown, that quantity of the consumed oxygen is proportional to the energy coming into the organism, whereas  the factor of proportionality (so-called oxi-caloric coefficient) is approximately equal to $4.821 \; cal/ml \; {\sf O}_2$ \ [19]. The estimating  of incoming energy by the speed of consumption of oxygen has received the name the method of indirect calorimetry.

 The energy of molecules of ATP is used basically for the synthesis of molecules, which are included in the  organism, also as for other processes requiring expenses of energy (muscle contraction, migration of cells, conduction of nervous impulses, etc.). Regarding the above-described classification of internal variables, the result of the release of energy of molecules of ATP is the emergence of internal variables of the second group, which is accompanied by excitation of internal variables of the third group. 

The flux of heat energy from the living  system $\Delta Q$ is measured by the method of  direct calorimetry, that is by direct measurement of the speed of heat production of the organism by means of a calorimeter. The amount of thermal energy coincides with the dissipation of energy that appears at the relaxation of internal variables of the third group with small times of  relaxation. The amount of the heat flux is evidence of the presence of quickly relaxating internal variables.

In steady state, when there are no changes of thermodynamic functions, incoming and outgoing flows of energy, according to equation (1),   are equal.  When an organism grows, the incoming stream of energy provokes the emergence of internal variables of the second and third group. The relaxation of variables of the third group defines dissipation of energy that in thermal form is deleted completely from the  organism. Remind  that the internal variables of the first and second groups do not contribute to the dissipation due to their huge relaxation times.  Thus, the difference between the incoming and outgoing flows, if exists, gives rise to the  emergence  of structural variables of the second group,   providing the  construction of organs and tissues of the organism.

Entropy $S$ of the whole organism,  due to the mass growth, increases;  simultaneously increases in the number of internal variables and their mutual  connections that is, one can say, increases in structural complexity of the organism. The better,  than the total entropy $S$,  characteristic of the living organism is the specific entropy $S/M$: under the balance of incoming and outgoing energy, specific entropy $S/M$, as it follows from equation (3),  can only decrease. In the situations of growth, the complexity of organism increases,  and, so as the greater the complexity, the less entropy, one can assume that the specific entropy does not increase,  and it gives
\begin{equation}
\frac{M}{S} \,\frac{\di}{\di t} \left(\frac{S}{M}\right) = - \sigma, \quad \sigma \geq 0.
\end{equation}

At the moment we cannot say that the quantity $\sigma$ is constant, but later (in Section 3.4), its value will be estimated and  appears to be about zero.

The permanent  exchange of substances and energy between the organism and environment, which  is a necessity for the creation of structures and maintenance of functioning, is controlled by the  program (the internal instructions) stored with internal variables of the first group.

\subsection{Equations of growth}

From the previous reasoning, it  follows that the growth rate of mass of the organism  is essentially determined by a difference between  quantities of incoming energy and outgoing heat. Equation (3) shows the linear dependence of the specific growth rate of mass on the difference, which had attracted  special attention and attained the name of $\Psi_u$-function [5]. The specific psi-function is defined as  
\begin{equation}
\psi_u = \frac{1}{M} \left(\frac{\Delta Q}{\Delta t} - \sum_{i} \, \Xi_{i} \,\frac{\di \xi_i }{\di t} + \sum_{\alpha}\, \eta_\alpha \, \frac{\Delta N_\alpha }{\Delta t} \right).
\end{equation}

Now one can reformulate relation (3), using definitions (5) and (6), so that the equation of growth can be recorded in the form 
\begin{equation}
\frac{1}{M} \, \frac{\di M}{\di t}  = \sigma + \kappa \,  \psi_u,
\end{equation}
where $\kappa  = M/TS$. According to the above equation,  the origin of growth is the difference  between quantities of incoming and outgoing flows of energy. The growth occurs until the organism will not reach the final steady state defined both by the genetic program, and conditions of the inhabitancy. The majority of poikilothermal animals grow during all life. The  homoiothermal animals  (mammals and birds) and some of the poikilothermal ones reach the stable mass [20]. 

In the stationary case the organism should be considered as a system,  in which all processes appear balanced:
\begin{equation}
\frac{\Delta Q}{\Delta t} - \sum_{i} \, \Xi_{i} \,\frac{\di \xi_i }{\di t} + \sum_{\alpha}\, \eta_\alpha \, \frac{\Delta N_\alpha }{\Delta t} =0.
\end{equation}
The adult organism, thus, can be considered as the certain constant structure that is being in the non-equilibrium stable state described constant thermodynamic functions. 

The time dependence of the mass growth for all types of animals can be approximated  by a simple equation  [21]:
\begin{equation}
\frac{\di M^{1/u}}{\di t}  = V_0 c^t, \quad 0 <c <1, 
\end{equation}
where $V_0$ is the speed of the growth in initial point in time $t=0$; $c$ is a parameter determining the  rate of growth; $u$ is a correction factor that is taking into account interfaced processes (morphogenesis,  differentiation, etc.).

The special case of equation (9) - the equation of Bertalanffy [22] -- can be recorded in the form 
\begin{equation}
\frac{1}{M} \, \frac{\di M}{\di t}  = \frac{u \ln c}{c^{t-t_0}- 1}, 
\end{equation}
where $t_0$ is a conditional time, at which assumingly $M = 0$, $c$ is a factor determining the rate of growth, $u$ is a factor connected with the influence of morphogenetic processes during  the growth.

\subsection{Testing of the thermodynamic equation of growth}

To estimate correspondence  of equation (7) to reality, we address to results of direct observation of growth of animals in the situations, when one can control the fluxes  of energy between the organism and environment [4, 5, 23 - 28].   The fluxes  of heat from the organism were measured by
 the method of  direct calorimetry. The fluxes of chemical energy in the organism were estimated by the method of indirect calorimetry according to the rate of oxygen consumption. We can draw certain conclusions from the results.

It has appeared, that during an essential period  of ontogenesis of organisms, the values of incoming and outgoing  energy fluxes practically coincide,
 \linebreak

\centerline{\includegraphics[scale=0.75]{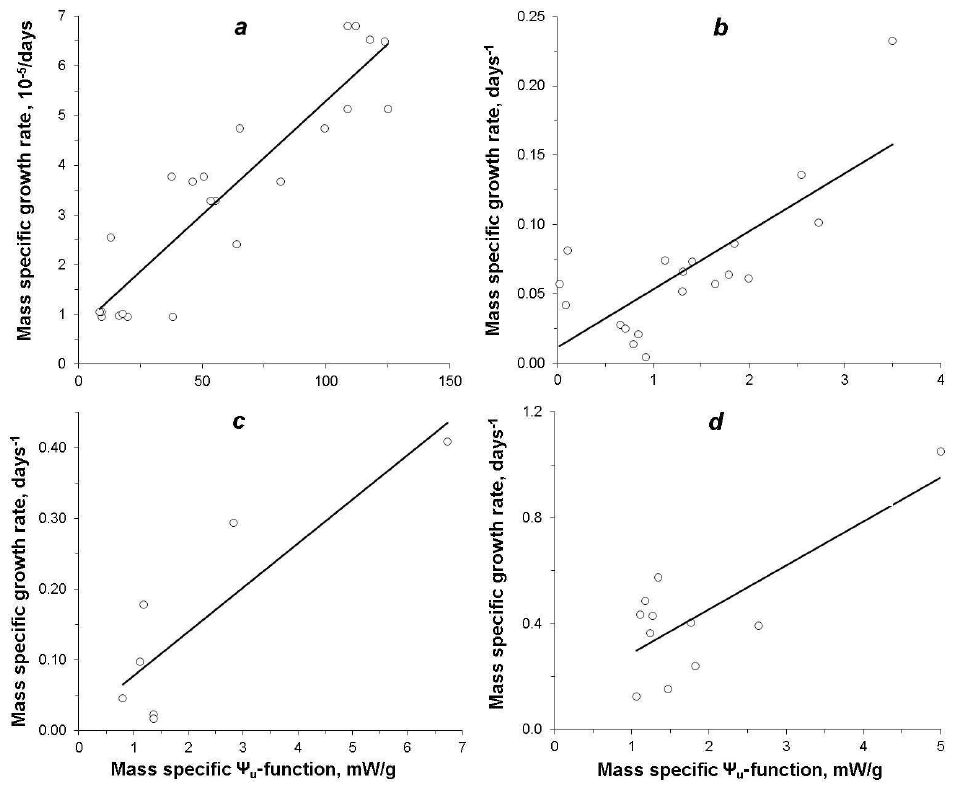}}

\vspace*{-90mm}

\centerline{\bf Figure 1 \ The dependence of specific growth rate on $\psi_u$-function} 

\noindent{\sl\small Species:  {\it a} - yeast {\it Saccharomyces cerevisiae} (according to Schaarschmidt et al. [24]; {\it b} - larvae of  crickets {\it Acheta domestica}; {\it c} - embryos of lizards {\it Lacerta agilis}; {\it d}  - chicken embryos {\it Gallus domesticus} (according to Kleymenov  [28]). The dependences  are approximated by straight lines, according to equation (7).
 
} 
\vspace*{3mm} 
\hrule 

\vspace*{3mm} 

\newpage

\begin{center}

{\bf Table of Estimates of Specific Entropy}\\

\vspace{4mm}
{\small
\begin{tabular}{|c|c|c|c|c|}

\hline
&&&&\\ 
Type (class) & $\kappa, \; g/mW\cdot day$ & $\sigma,\; day^{-1} $ & $r^2$& \parbox[c]{12mm}{$\; S/M$, \\  $J/g\cdot K$}
 \\ &&&& \\

\hline \hline &&&&\\

\parbox[c]{28mm}{{\it Saccharomyces cerevisiae}
(Saccharomycetes)$^1$} &  $(4.73\pm 0.43) \cdot 10^{-7}$ &	$(5.70\pm 2.98)\cdot10^{-6}$	&$0.92 \pm 0.08$&	$365\cdot10^6$ \\ 
&&&& \\

\parbox[t]{28mm}{{\it Acheta domestica } (Insecta), larvae$^2$} & $0.042 \pm 0.009$&	$0.011 \pm 0.014$&$	0.76 \pm 0.16	$&4400
\\ &&&& \\

\parbox[t]{29mm}{{\it Lacerta agilis $\;\;\;\;\;$ } (Reptilia),\,embryos$^2$}& 0.062 $\pm 0.016$&	$0.015 \pm 0.046$&	$0.87 \pm 0.22$&	3350
\\ &&&& \\

\parbox[t]{28mm}{{\it Gallus domesticus} (Aves), embryos$^2$} & $0.166 \pm 0.046$	&$0.123 \pm 0.098$&$	0.98 \pm 0.07	$&300\\ &&&& \\
\hline 

\end{tabular}}

\vspace*{3mm}
{\small Sources of data: 1 -- Schaarschmidt et al. [24]; 2 - Kleymenov [28]}.

\end{center}
\vspace*{3mm}
\hrule 
\vspace*{4mm}
\noindent
and such a situation has received the name of the  {\it current steady state} of the  organism  [4, 7, 18]. However, there is mass growth in such situations, and, according to equation (7), the difference  between incoming and outgoing flows of energy must exist. One can assume that the  growth occurs so slowly, that the difference appears to be so small, that, apparently, cannot be detected with the used methods. Apart of it, when the method of indirect calorimetry is used for  the assessment of incoming stream of energy, the contribution of processes of glycolysis, which are also processes  of synthesis of ATP, are neglected in comparison with the process of oxidizing phosphorylation. However, the contribution of the processes of glycolysis  can appear comparatively essential in this phase of growth.  

The authentic distinction between data of the indirect (according to the  rate of oxygen consumption) and the direct (according to the  rate of heat production) calorimetry in situations of the constant environment is 
observed in cases of the initialization of such processes, as differentiation, morphogenesis etc [28]. At early stages of the growth in ontogenesis, the  distinction of results of the direct and indirect calorimetry $\psi_u$ is great enough to address to an assessment of the adequacy of equation (7).


\centerline{\includegraphics[scale=0.75]{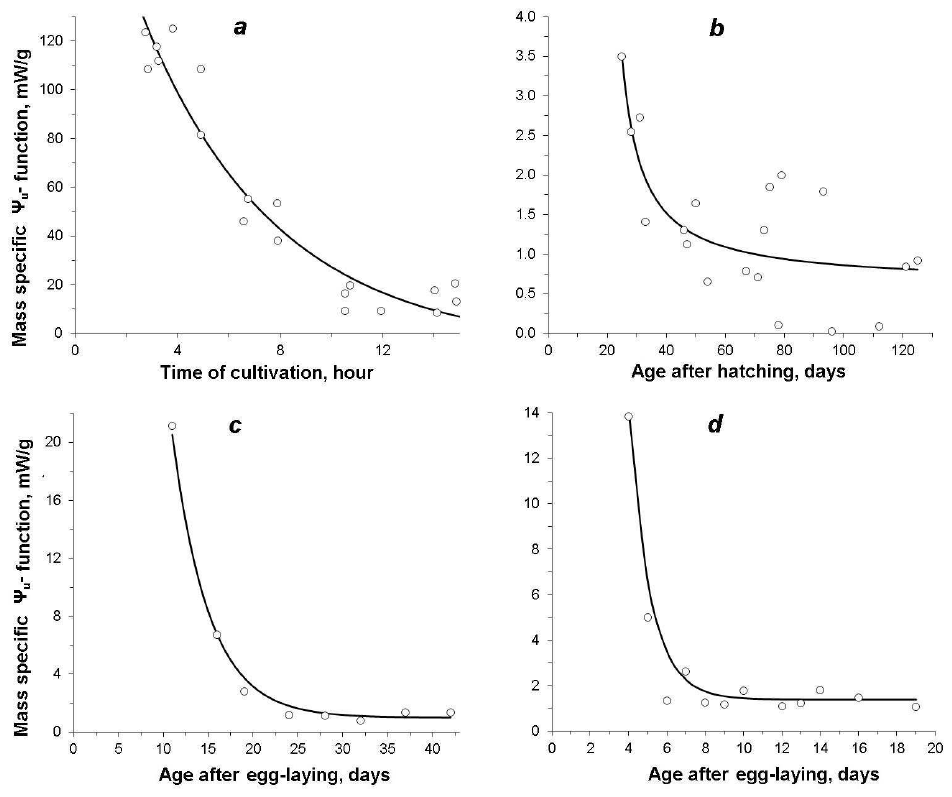}}

\vspace*{-90mm}

\centerline{\bf Figure 2 \ The time dependence of $\psi_u$}

\noindent{\sl\small Species:  {\it a} - yeast {\it Saccharomyces cerevisiae} (according to Schaarschmidt et al.  [24]); {\it b} - larvae of  crickets {\it Acheta domestica}; {\it c} - embryos of lizards {\it Lacerta agilis}; {\it d}  - chicken embryos {\it Gallus domesticus} (according to Kleymenov [28]). The dependences  are approximated by curve lines, according to equation (11).
} 
\\

\hrule 

\vspace*{3mm}
\newpage 
The specific growth rates, as functions of $\psi_u$,  according to experimental data [24, 28] are shown on Fig.\,1. The values of coefficients  of correlation and parameters of equation (7) are collected in Table. Though the variability of data is great, the coefficient of correlation between the $\psi_u$ and specific growth rate differ reliably from zero ($p <0.01$). Note that the quantity $\sigma$ is not significantly  different from zero considering the error bars.  Thus, specific entropy can be  considered  as a constant during a significant time interval, and the calculated values of the coefficient $\kappa$ allow to estimate specific entropy $S/M = 1/\kappa \, T$ (see the Table). It is seen that specific entropy 
decreases in the evolutionary raw: yeast $\rightarrow$ insects $\rightarrow$ reptiles $\rightarrow$ birds. These results show a reduction of entropy or, in other words,  an increase of a degree of complexity of living systems during evolution.

A positive correlation between the specific growth rate and $\psi_u$ also has been noted for representatives of different taxons: bacteria {\it Serratta marinorubta, S. mercescens} [27], yeast {\it Sacharamyces cerevisiae} [24],  larvae of mealworm beetle {\it Tenebrio molitor} [25], larvae of house crickets {\it Acheta domestica}, embryos of sand lizard {\it Lacerta agilis}, chicken embryos {\it Gallus domesticus} [28], juvenile fishes {\it Xiphophorus helleri} [23], larvae of  axolotl {\it Ambystoma mexicanum} [26], ovogenesis of clawed frogs {\it Xenopus laevis} [5].

It is important to have a  look  at the  time dependence  of the quantity $\psi_u$ at the growth of the organisms: it gives an additional way of assessment of adequacy of the equation (7).  Comparing equation (10) with the equation (7), we receive:
\begin{equation}\psi_u  = \frac{a}{1-c^{t-t_0}} - b, 
\end{equation}
 where $a = -u \ln c/\kappa  ; b = \sigma/\kappa  $. On the plots of Fig.\,2, the experimental values of $psi$-function,  corresponding to values in Fig.\.1, are shown with empty circles. The solid curves are calculated, according to equation (11), at individual values of quantities  $a$, $b$  and  $c$.  This comparison confirms  the existence of the universal law of growth in the form (10).

\section{The  Discussion and Conclusions}

The thermodynamic approach for the open systems, developed by Prigogine [9] and his followers, gives a general foundation for description of dynamics of  processes running in living  systems. The internal variables,  the number of which is huge, describe the deviations of the biological organism,  as a thermodynamic system, from equilibrium and appears to be the most  essential elements of description. However,  we cannot even inventory all internal variables, and are still very distant from the detailed  description of the living  organism. We have found that, though the biological organism exists as integral system and all internal variables are anyhow connected with each other, it is useful  to allocate  some groups of the internal variables according to the relaxation times.   The mechanisms of heredity and control, which is the main thing that distinguish the biological organism  from a system of coupled chemical reactions, is assumed to be ascribed to a special group (the first group in our classification) of internal variables. The internal variables with relaxation times comparable with the lifespan of organism (the second group) describe  the metastable  structure that is a real implementation of living organism.  The variables of the third group are participating in  thermodynamic processes that are obeying the linear laws of the  non-equilibrium  thermodynamics. 

The clarification of functional roles of the internal variables allows us to formulate the  thermodynamic equation of growth of organisms that corresponds not only to the results of experimental observations but also to basic phenomenological equations of growth. It is remarkable that application of the derived equation of growth to experimental data allows one to estimate the value of specific entropy of living organisms. 

The specific entropy appears to be the most important thermodynamic characteristic of the living organism. In some examples, it was shown that  specific entropy remains constant during the growth and development of the organism. It is possible that this property is valid only for many-cell bodies, when the growth is connected with the  multiplication of the high complexity of the original cell. So as the specific entropy does not change substantially during the lifespan, it appears to be an individual characteristic of the organism.  The assessments of specific entropy for larvae of crickets, embryos of the sand lizard and chicken embryos confirm that there is  decrease of specific entropy and, consequently, an increase of a degree of complexity of the living  systems during progressive macroevolution.

\setcounter{equation}{0}
\renewcommand{\theequation}{A.\arabic{equation}}

\section*{Appendix A: Entropy of the open thermodynamic system }

The concept of entropy, as a measure of the degraded energy, has arisen at the description of the ideal equilibrium situations. Variation of entropy $S$ under reversible process of a heat  transaction to system in the amount $\Delta Q$ is defined as 
\begin{equation}
\Delta S = \frac{\Delta Q}{T}, 
\end{equation}
This circumstance allows to interpret entropy of the systems in equilibrium situations  as a measure of thermal, completely degraded energy under some restrictions by a set of constitutive  parameters (volume $V$ in the simplest case) that are defined the thermodynamic system.  At the extension of concept of entropy for non-equilibrium states of the system,  it is natural to demand, that interpretation of entropy as a measure of thermal, completely degraded energy is being kept. The ignoring of this requirement has brought to existence of a set of  'non-equilibrium entropies' and appears to be a source of some misinterpretation.  

We shall demonstrate the realization of such an approach, following the works [13, 14],   for general case of  open system, which is assumed performs no work. Then, the  expression for variation of total internal energy $U$ (the first law of thermodynamics) is recorded in the form of 
\begin{equation}
\Delta U = \Delta Q + \sum_{\alpha} \, (\eta_\alpha + \mu_\alpha) \, \Delta N_\alpha, 
\end{equation}
where $\Delta Q$ is a flux of heat into the system, $\Delta N_\alpha $ is a change of substance of a kind $\alpha$ within  the  system. The enthalpy of unity of the substance of  kind  $\alpha$ is presented as a sum of two quantities, $\eta_\alpha + \mu_\alpha$,   one of them $\mu_\alpha$ is chemical potential of the substance. 

The parity (A.2) is valid both for equilibrium, and for non-equilibrium situations, however, if, in equilibrium states, internal energy $U$ represents completely degraded energy,  in a non-equilibrium situation only some part of the total internal energy $U$, which we shall designate as $E$, is in the form of thermal energy. Other part of energy is connected with the restrictions caused by constant constitutive parameters (volume $V$ in the simplest  case) and relaxating internal parameters $\xi$. The latter are introduced to describe deviations of the  thermodynamic system from the equilibrium state. The total internal energy of the system changes due to the change of the composition of the system\footnote{Thanks to W. Muschik, who pointed at the  existence of this contribution.} and  work on variations of the internal parameters $\xi$,  so that it can be  recorded as 
  \begin{equation}
\Delta U = \Delta E + \sum_{j} \, \Xi_{j} \,\Delta \xi_j + \sum_{\alpha} \, \mu_\alpha \, \Delta N_\alpha. 
\end{equation}
The part of total internal energy is in the thermal form, and a part in the form that is ready to turn in thermal energy.

There are various ways of definition of entropy  for the open thermodynamic system in view of all of non-equilibrium processes. In the simplest  version, one  keeps  definition (A.1), which, apparently, is wrong, because since times of Clausius is known that in non-equilibrium processes for the closed system 
 \begin{equation}
\Delta S \geq \frac{\Delta Q}{T}
\end{equation}
In view of it, the definition of non-equilibrium entropy for the closed system should contain additional positive terms
$$
\Delta S = \frac{\Delta Q}{T} + ....
$$

To define entropy as a measure of thermal energy, we shall return to the law of conservation and transformation of energy for the system, which can be recorded, according to   parities (A.2) and (A.3)  in the form of 
\begin{equation}
\Delta E = \Delta Q - \sum_{i} \, \Xi_{i} \,\Delta \xi_i 
+ \sum_{\alpha} \, \eta_\alpha \, \Delta N_\alpha. 
\end{equation}
Further, adhering to interpretation of entropy as a measure of energy in the form of heat, we follow an equation (A.1) and, neglecting the work of the system, record 
\begin{equation}
\Delta S = \frac{\Delta E}{T}.
\end{equation}
Here temperature $T$ is also defined as absolute temperature for the thermal part  of the system that has  thermal energy $E$. Of course, the real temperature of the system in a non-equilibrium state could differ from this temperature $T$, but the difference can be considered as an internal variable and included in the list of $\bxi $(inexplicitly).  

In such a way, equation (A.5) and (A.6) determines the  differential of entropy 
\begin{eqnarray}
dS &=& \frac{1}{T} \, \left(\Delta Q - \sum_{i} \, \Xi_{i} \,\Delta \xi_i 
 + \sum_{\alpha} \, \eta_\alpha \, \Delta N_\alpha \right).
\end{eqnarray}

Equation (A.7) is the only definition of entropy of the open system, which is consistent with the understanding this quantity as a measure of the degraded energy and empirical evidences.

\newpage

\section*{References}

\begin{enumerate}

{\small

\item E.S. Bauer, Theoretical Biology, All-Union Inst. Exp. Med., Moscow-Leningrad, 1935. 

\item I. Prigogine, J.M. Wiame,  Biologie et thermodynamique des phenomenes irreversibles, Expenentia  2 (1946),  451-453.

\item H.E. Salzer, Toward a Glbbsian approach to the problems of growth and cancer, Acta Biotheor. 12 (1957), 135-166.

\item A.I. Zotin,  {\it Thermodynamic aspects of developmental biology,} Monographs in Developmental Biology. V. 5, S. Karher,  Basel et al., 1972. 159 p.

\item A.I. Zotin,   Dissipative structures and $\Psi_u$-function,  In: {\it Thermodynamics of Biological Processes, } Walter de Gruyter, Berlin \& NY, 1978,  P. 301-304. 

\item A.I. Zotin,  Changes of $\psi_d$ and $\psi_u$-functions during oogenesis of Xenopus laevis, In  {\it Thermodynamics of Biological Processes,} Walter de Gruyter, 1978,  P. 191-196.

\item A.A. Zotin, A.I.  Zotin,  Phenomenological theory of ontogenesis,  {\it Int. J. Dev. Biol.} {\bf 41}, (1997),  917-921.

\item E.  Schrodinger,  {\it What is Life? The Physical Aspect of the Living Cell}, Cambridge Univ. Press, Cambridge, 1944.

\item I. Prigogine,  {\it Introduction to Thermodynamics of Irreversible Processes. Second Revised Edition,.} Interscience publ., N.Y.-London,  1961.

\item K.S. Trincher, [{\it Biology and information. Elements of biological thermodynamics}], Nauka, Moscow, 1965,  119 p.

\item G. Nicolis and I. Prigogine,  {\it Self-Organisation in Non-Equilibrium Systems: From Dissipative Structures to Order through Fluctuations}, John Wiley \& Sons, New York, 1977. 

\item  W.B. Cannon, The Wisdom of the Body, W.W. Norton \& Company, inc., London, 1932.

\item V.N. Pokrovskii,  Extended thermodynamics in a discrete-system approach, {\it Eur. J. Phys.} {\bf 26}, (2005), 769-781.

\item V.N. Pokrovskii,  A derivation of the main relations of non-equilibrium thermodynamics. Hindawi Publishing Corporation: {\it ISRN Thermodynamics}  {\bf 2013}, article ID 906136, (2013),  9 p.  http://dx.doi.org/10.1155/2013/906136 

\item L. Brillouin {\it Science and information theory}. Academic Press Inc. Publishers, New York, 1956. 

\item P.D. Gujrati, Hierarchy of Relaxation Times and Residual Entropy: A Nonequilibrium Approach, {\it   Entropy} {\bf  20} (2018), 149, doi:10.3390/e20030149

\item D. Metzler, {\it Biochemistry: the chemical reactions of living cells,} Academic Press, London et al., 1977, 1168 p.

\item A.I. Zotin, R.S. Zotina,  [{\it Phenomenological theory of development, growth and aging of organisms}],  Nauka, Moscow, 1993,  363 p.

\item D.Briedis, R.C. Seagrave,  Energy transformation and entropy production in living systems. I. Application to embryonic growth,  {\it J. Theor. Biol.} {\bf 110}  (1984), 173-193.

\item A.A. Zotin,  [{\it Patterns of growth and energy metabolism in the ontogeny of mollusks}], Doctor dissertation thesis. Moscow: IBR RAN, (2009), 334 p.

\item A.A. Zotin, The united equation of animal growth,  {\it Amer. J. Life Sci.} {\bf 3} (2015), 345-351.

\item L. von Bertalanffy,   On the von Bertalanffy growth curve,  {\it Growth} {\bf 30} (1965), 123-124.

\item V.A. Grudnickij,  [{\it Growth and metabolism in swordtails}]. PhD thesis. Moscow: IBR RAN, (1972), 137 p.

\item B. Schaarschmidt, A.I. Zotin, R. Brettel, I. Lamprecht, Experimental investigation of the bound dissipation function: change of the (u-function during growth of yeast,  {\it Arch. Microbiol.} {\bf 105} (1975),  13-16.

\item K.D. Loehr, P. Sayyadi, I. Lamprecht,  Thermodynamic aspects of development for Tenebrio molitor L,  {\it Experientia} {\bf 32} (1976),  1992-1003.

\item V.A. Grudnickij, I.S. Nikol'skaja,   [Heat production at early stages of axolotl growth from direct and indirect calorimetry data], {\it Ontogenesis} {\bf 8} (1977), 80-82.

\item J. Bermudez, J. Wagensbers,  The entropy production in microbiological stationary states,  {\it J. Theor. Biol.} {\bf 1986} (1986) , 347-358. 

\item S.Ju. Kleymenov,   {\it The mass specific energy metabolism in the early ontogenesis of animals from the data of indirect and direct calorimetry,} PhD thesis. Moscow: IBR RAN, (1996), 147 p.
}

\end{enumerate}

\setcounter{equation}{0}
\renewcommand{\theequation}{B.\arabic{equation}}

\section*{ Appendix  B. \  Correction \  to the paper \  {\it The growth and development of living organisms from the thermodynamic point of view} }
\vspace{5mm}

Our derivation of the thermodynamic equation of growth and development of living organisms [1] was based on the equation of the entropy change in the  open system, that is, on the equation
\begin{equation}
T\,dS = \Delta Q - \sum_{j} \, \Xi_{j} \,\Delta \xi_j  + \sum_{\alpha =1}^k\, \mu_\alpha \, \Delta N_\alpha. 
\end{equation}
The last term in this equation was identified  with total energy that is coming into the system with  streams of  substance through borders,  where $\mu_\alpha$ was understood mistakenly  as a total specific energy of a molecule of kind $\alpha$ in the system; the other symbols are explained earlier [1].  This  appears to be  incorrect: the specific total energy, coming with the streams of substances,  ought to be written as the sum of chemical potential $\mu_\alpha$ and  specific thermal energy $\eta_\alpha$, and only the second part of it  gives contribution to change of entropy\footnote{Note that equations  (B.1)   reflects  the Prigogine's  description  of the open system  [2] and corresponds to his equation (3.47); to make this clear, we can recall that  the auxiliary quantity  $\Phi$ in equation (3.47) is  understood by Prigogine [2, equation (2.14)] as the total amount of energy, coming into the system,
$$
d\Phi = dQ + \sum_{\alpha}\, (\mu_\alpha+\eta_\alpha) \, d_e n _\alpha,
$$  
 where $\mu_\alpha+\eta_\alpha$ is total specific energy, coming into the system with the  streams of  molecules,  so that Prigogine's equation (3.47) can be rewritten as 
$$
T\,dS = d Q +A \, d \xi_j + \sum_{\alpha}\, \eta_\alpha \, d_e n_\alpha   
$$
In contrast to this correct equation, the last term of our equation (B.1) was understood as total specific energy, which can be written in the conventional terms as the sum of  chemical potential and  specific thermal energy,  $\mu_\alpha + \eta_\alpha$. Equations (B.2) and (B.3) give a correct description of the situation.}.  The situation could be conveniently described [3] with help of the two thermodynamic functions:  Gibbs free energy and  entropy of the system, respectively,
\begin{eqnarray}
dG &=&  \sum_{j} \, \Xi_{j} \,\Delta \xi_j+ \sum_{\alpha} \, \mu_\alpha \, \Delta N_\alpha, \\  
T\,dS &=& \Delta Q - \sum_{j} \, \Xi_{j} \,\Delta \xi_j  + \sum_{\alpha}\, \eta_\alpha \, \Delta N_\alpha, 
\end{eqnarray}
The last terms in equations (B.2) and (B.3) represent the flows of chemical energy incoming or outgoing with the stream of molecules of substances  $\Delta N_\alpha$ through the borders of the system. The quantities  $\Xi_{j}$ in equations (B.2) and (B.3)  are   thermodynamic forces, corresponding to internal variables  $ \xi_j $.  For a chemical process, the internal variable  is determined as incompleteness of chemical reaction and the corresponding thermodynamic force --  as  affinity of the reaction with the  negative sign, $\Xi_{j} = -A_j$. In addition to chemical reactions, equations (B.2) and (B.3)  may include various variables describing processes  of leveling of temperature and concentrations of substances. 

Equations (B.2) and (B.3)  can be considered as a proper generalization of the Prigogine's principles of description  of the open system; in particular, equation (B.3) is a correct generalization of equation (3.47) from the book by Prigogine [2].  

The further arguments in the article [1] seem to be correct, and the growth equation (7) retains its previous form 
\begin{equation}
\frac{1}{M} \, \frac{\di M}{\di t}  = \sigma + \kappa \,  \psi_u,
\end{equation}
while the coefficient changes to $\kappa  = M/(G+TS)$. The specific psi-function instead of (6) in [1]  is defined now as  
\begin{equation}
\psi_u = \frac{1}{M} \left(\frac{\Delta Q}{\Delta t} - \sum_{i} \,( A_j  + \Xi_{i})\,\frac{\di \xi_i }{\di t} + \sum_{\alpha=1}^K\, (\mu_\alpha+\eta_\alpha) \, \frac{\Delta N_\alpha }{\Delta t} \right), 
\end{equation}
while the middle term on the right side of this equation can be omitted.
 
One can see that the written  equations of growth and development  (B.4) and (B.5) for $G=0$ coincide with equations (6) and (7) in [1]. 

 It is easy to see  that estimates of specific entropy $S/M$ of the living organisms in Table 1 in [1] are, taking the correction into account,  estimates of the upper absolute value of the specific entropy. The main conclusion of the study remains unchanged:  the complexity of a living object can be related to the value of the specific entropy $S/M$: the lower the specific entropy, the greater the structural complexity of the organism.

Unfortunately,   Alexei A. Zotin  was not be able to read this compendium;  he died 21.05.2021   after a serious illness, without completing all   his projects.      The essence of the correction  was introduced into the  refined version of the derivation of thermodynamic equation of growth and development of living organisms, which  has appeared \ in Chapter 9 of the monograph~[3].

\vspace{5mm}

\centerline{References} 

\begin{enumerate}

{\small 

\item A.A. Zotin, V.N. Pokrovskii,  The growth and development of living organisms from the thermodynamic point of view,  Physica A: Statistical Mechanics and its Applications 512 (2018), 359 - 366.

\item I. Prigogine, Introduction to Thermodynamics of Irreversible Processes.  Third Edition, Wiley \& Sons, 1967.

\item V.N. Pokrovskii,  Thermodynamics of Complex Systems: Principles and  applications, IOP Publishing, Bristol, UK, 2020.  
}
\end{enumerate}
\end{document}